\documentclass[]{aastex631}

\newcommand{\omegagw}{\Omega_{\rm{gw}}}
\newcommand{\baromega}{\bar{\Omega}_{\rm{gw}}}
\newcommand{\mc}{\mathcal{M}_{c}}
\newcommand{\fobs}{f_{\rm{o}}}
\newcommand{\eobs}{\hat{e}_{\rm{o}}}

\begin{document}

\title{Searching for anisotropic stochastic gravitational-wave backgrounds \\with constellations of space-based interferometers}


\author[0000-0003-0889-1015]{Giulia Capurri}
\affiliation{SISSA, via Bonomea 265, 34136, Trieste, Italy}
\affiliation{INFN-Sezione di Trieste, via Valerio 2, 34127 Trieste, Italy}
\affiliation{IFPU, via Beirut 2, 34151, Trieste, Italy}

\author[0000-0002-4882-1735]{Andrea Lapi}
\affiliation{SISSA, via Bonomea 265, 34136, Trieste, Italy}
\affiliation{INFN-Sezione di Trieste, via Valerio 2, 34127 Trieste, Italy}
\affiliation{IFPU, via Beirut 2, 34151, Trieste, Italy}
\affiliation{IRa-INAF, Via Gobetti 101, 40129 Bologna, Italy}

\author[0000-0003-3127-922X]{Lumen Boco}
\affiliation{SISSA, via Bonomea 265, 34136, Trieste, Italy}
\affiliation{INFN-Sezione di Trieste, via Valerio 2, 34127 Trieste, Italy}
\affiliation{IFPU, via Beirut 2, 34151, Trieste, Italy}

\author[0000-0002-8211-1630]{Carlo Baccigalupi}
\affiliation{SISSA, via Bonomea 265, 34136, Trieste, Italy}
\affiliation{INFN-Sezione di Trieste, via Valerio 2, 34127 Trieste, Italy}
\affiliation{IFPU, via Beirut 2, 34151, Trieste, Italy}

\begin{abstract}

Many recent works have shown that the angular resolution of ground-based detectors is too poor to characterize the anisotropies of the stochastic gravitational-wave background (SGWB). For this reason, we asked ourselves if a constellation of space-based instruments could be more suitable. We consider the Laser Interferometer Space Antenna (LISA), a constellation of multiple LISA-like clusters, and the Deci-hertz Interferometer Gravitational-wave Observatory (DECIGO). Specifically, we test whether these detector constellations can probe the anisotropies of the SGWB. For this scope, we considered the SGWB produced by two astrophysical sources: merging compact binaries and a recently proposed scenario for massive black-hole seed formation through multiple mergers of stellar remnants. We find that measuring the angular power spectrum of the SGWB anisotropies is almost unattainable. However, it turns out that it could be possible to probe the SGWB anisotropies through cross-correlation with the CMB fluctuations. In particular, we find that a constellation of two LISA-like detectors and CMB-S4 can marginally constrain the cross-correlation between the CMB lensing convergence and the SGWB produced by the black hole seed formation process. Moreover, we find that DECIGO can probe the cross-correlation between the CMB lensing and the SGWB from merging compact binaries.
\end{abstract}

\keywords{Gravitational wave astronomy(675) --- Gravitational wave detectors(676) --- Gravitational wave sources(677) --- Cosmic anisotropy(316)}

\section{Introduction} \label{sec:intro}

The stochastic gravitational-wave background (SGWB) is a diffuse gravitational-wave signal resulting from the superposition of numerous unresolved sources \citep{2019RPPh...82a6903C}. The scientific community usually classifies the SGWB into two categories according to its origin. On the one hand, there is the cosmological SGWB, produced during the early phases of the Universe by many possible sources (e.g., inflation, reheating, pre-Big Bang scenarios, cosmic strings, and phase transitions). On the other hand, gravitational-wave sources that have been active since the beginning of the stellar activity (e.g., compact binaries, intermediate/extreme mass ratio inspirals, rotating neutron stars, and core-collapse supernovae) give birth to the astrophysical part of the SGWB. 
Because of the information richness that it encodes, the characterization of the SGWB is one of the main targets of present and future gravitational wave detectors. Existing data from the LIGO/Virgo/KAGRA \citep{2015CQGra..32g4001L,2015CQGra..32b4001A,2012CQGra..29l4007S} network have already placed upper bounds on both the isotropic and anisotropic components of the SGWB \citep{2021PhRvD.104b2004A,2021PhRvD.104b2005A}.

A possible way to disentangle the various components of the SGWB is to study the frequency dependence of its amplitude: different backgrounds, indeed, display different frequency spectra. Like other cosmic backgrounds, the SGWB is mainly isotropic with a tiny anisotropic component. The SGWB anisotropies are another possible tool to distinguish among the different components. Moreover, they constitute a tracer of the large-scale distribution of matter in the Universe. The anisotropies of the cosmological SGWB reflect the matter distribution in the early Universe \citep{2019PhRvD.100l1501B,2020JCAP...02..028B,2021PhRvD.103b3522V}. Instead, the sources of the astrophysical SGWB typically reside inside galaxies, whose distribution traces the Large Scale Structure \citep{2021JCAP...02..035L,2018JCAP...09..039S}. 

Among the multiple sources of SGWB, the one given by the incoherent superposition of gravitational-wave events produced by merging stellar remnant compact binaries has raised significant interest among the scientific community \citep{2011RAA....11..369R,2011PhRvD..84h4004R,2011PhRvD..84l4037M,2011ApJ...739...86Z,2013MNRAS.431..882Z,2012PhRvD..85j4024W,2015A&A...574A..58K,2016PhRvL.116m1102A,2018PhRvL.120i1101A,2021PhRvD.103d3002P}. This astrophysical SGWB is expected to be the dominant component in the frequency range explored by ground-based detectors (Hz-kHz) and is likely to be the first to be detected. During the last few years, many studies focused on this anisotropic SGWB, with relevant effort given to theoretical modeling \citep{2017PhLB..771....9C,2017PhRvD..96j3019C,2018PhRvD..97l3527C,2018PhRvD..98f3501J,2019PhRvL.122k1101J,2018PhRvL.120w1101C,2019PhRvD.100f3004C,2020PhRvD.101j3513B,2020PhRvD.101h1301P,2020MNRAS.493L...1C,2021JCAP...11..032C,2022JCAP...06..030B}, observational searches \citep{2022PhRvD.105l2001A,2021PhRvD.104b2005A,2022Galax..10...34R,2022PhRvD.105b3519R,2022JCAP...11..009B,2021JCAP...03..080M,2020PhRvD.102d3502C,2019PhRvL.122h1102R}, and data analysis techniques \citep{2009PhRvD..80l2002T,2014PhRvD..90h2001G,2015PhRvD..92d2003R,2018PhRvD..98b4001A,2018MNRAS.481.4650R,2019MNRAS.487..562C,2013PhRvD..88h4001T}.

The two main obstacles to the detection of the SGWB anisotropies are the poor angular resolution of gravitational-wave detectors to a diffuse SGWB mapping and the presence of a considerable shot noise contribution. The former issue is mainly related to the noise properties of the detector and how they project onto the sky. However, it also depends on the network configuration and the scan strategy \citep{2020PhRvD.101l4048A}. On the other side, the shot noise arises because the SGWB is composed of discrete transient events occurring at a relatively low rate. Several recent studies have addressed the issue of shot noise, showing that its power spectrum is orders of magnitude higher than the intrinsic SGWB correlation induced by the Large Scale Structure \citep{2019PhRvD.100h3501J,2019PhRvD.100f3508J,2020PhRvD.102b3002A,2021JCAP...11..032C}. Cross-correlation with other tracers of the Large Scale Structure has already been proposed as a possible solution to reduce the impact of shot noise. In particular, several studies explored the potential of cross-correlations with galaxy number counts \citep{2020PhRvD.102b3002A,2020PhRvD.102d3513C,2021MNRAS.500.1666Y,2020MNRAS.491.4690M} and Cosmic Microwave Background (CMB) temperature fluctuations \citep{2021PhRvL.127A1301R,2021PhRvD.104l3547B,2022Univ....8..160C}. 

Many of those works have shown that the poor angular resolution of ground-based instruments prevents measuring the angular power spectrum of the SGWB anisotropies. It turned out that the cross-correlation with other probes can reduce the impact of instrumental and shot noise, but unfortunately not enough to guarantee detection. Therefore, we asked ourselves if a constellation of space-based detectors could be more suitable for the purpose. Indeed, the increased distance among the instruments constitutes a long interferometric baseline that should lead to a better angular resolution. In particular, we focused on two instruments: the Laser Interferometer Space Antenna (LISA) \citep{2022GReGr..54....3A,2017arXiv170200786A,2013arXiv1305.5720E} and the Deci-hertz Interferometer Gravitational-wave Observatory (DECIGO) \citep {2021PTEP.2021eA105K,2017JPhCS.840a2010S,2011CQGra..28i4011K}. The former is an ESA/NASA mission probably operative after 2034. It will be composed of three spacecraft in orbit around the Sun, operating as a correlated set of three laser interferometers searching for gravitational waves around the mHz frequency regime. The latter instrument, instead, is a Japanese proposed space gravitational wave antenna, targeting deci-Hz
gravitational wave signals. DECIGO is by itself a constellation of four clusters, each of which is composed of three spacecraft. Moreover, we also investigate the benefits of using a constellation of multiple LISA-like cluster

As a case study, we considered two different astrophysical sources of SGWB at these frequencies. On the one side, stellar compact objects binaries \citep{2019ApJ...881..157B,2021ApJ...907..110B}; on the other side, intermediate and extreme mass ratio black hole binaries formed in the center of dusty star-forming galaxies through the migration induced by dynamical friction with the gaseous environment, constituting a possible formation process of massive black hole seeds \citep{2020ApJ...891...94B,2021JCAP...10..035B}. 

The plan of the paper is as follows. In Section \ref{sec:detectors}, we describe detectors, with a particular interest in our treatment of their noise properties and the computation of the angular sensitivity. In Section \ref{sec:sources}, we review the processes that produce the SGWB we considered for this work. In Section \ref{sec:monopole}, we discuss the detection prospects for the isotropic amplitude of the SGWB. In Section \ref{sec:anisotropies}, we present the results of the SGWB anisotropies. In particular, we compute their amplitude and discuss the detection prospects. In Section \ref{sec:cross}, we show the results concerning the cross-correlation between the SGWB and the CMB lensing. Finally, in Section \ref{sec:conclusion}, we wrap up with some final considerations.

Throughout this work we adopt the standard flat $\Lambda$CDM cosmology with parameter values from the Planck 2018 legacy release \cite{2020A&A...641A...6P}, with Hubble rate today corresponding to $H_{0}= 67.4$ km s$^{-1}$ Mpc$^{-1}$, Cold Dark Matter (CDM) and baryon abundances with respect to the critical density corresponding to $\Omega_{\rm{CDM}}h^{2}= 0.120$ and $\Omega_{b}h^{2}= 0.022$, respectively, reionization optical depth $\tau=0.054$, amplitude and spectral index of primordial scalar perturbations corresponding to ln$(10^{10}A_{S})=3.045$ and $n_{S}=0.965$, respectively.

\vspace{1.5em}
\section{Detectors} \label{sec:detectors}

\begin{figure*}
\plotone{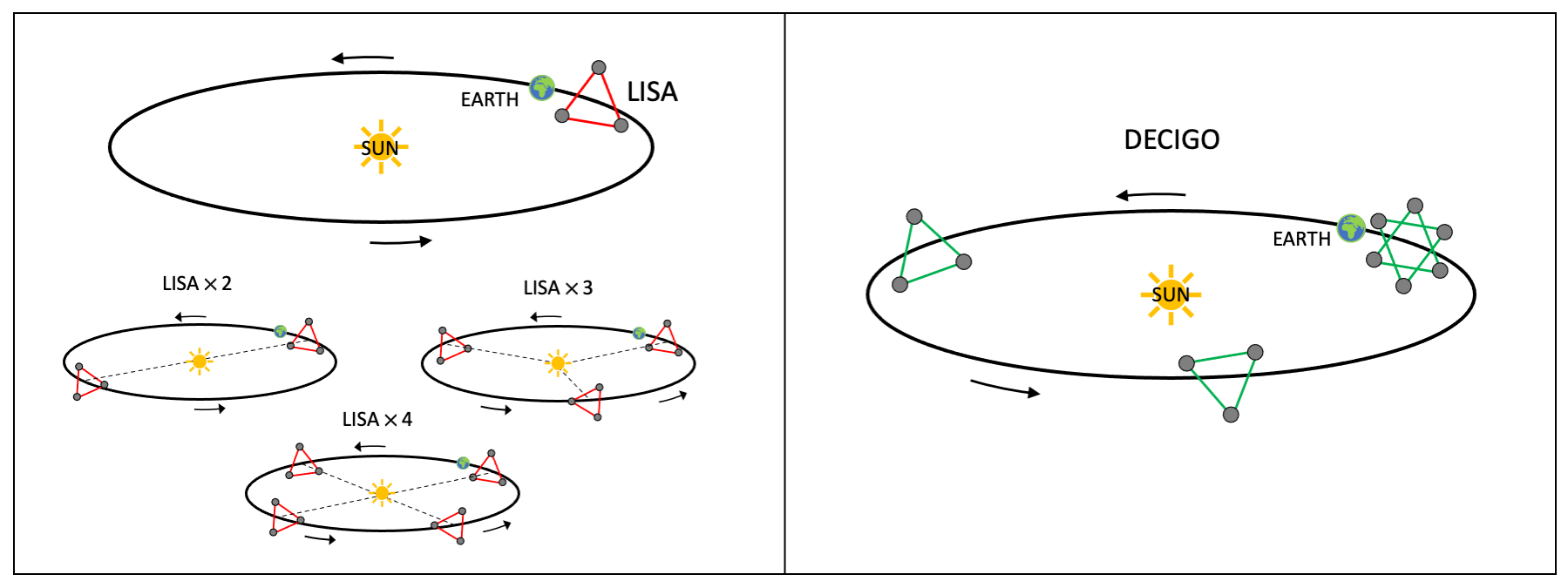}
\caption{Cartoon representation of the proposed configurations for LISA (left) and DECIGO (right), described in the text. We also show the three different constellations of LISA-like clusters we use for this work.
\label{fig:lisa_decigo}}
\end{figure*}

In this Section, we provide a general overview of the instruments we consider in this work: the Laser Interferometer Space Antenna (LISA) and the Deci-hertz Interferometer Gravitational-wave Observatory (DECIGO). After a brief description of the main characteristics of the detectors, we focus on a more detailed report of the specific prescriptions for the noise curves and the angular sensitivity we used to produce our results.

\vspace{.5em}
\subsection{LISA}

LISA \citep{2022GReGr..54....3A,2017arXiv170200786A} is a space-based gravitational wave observatory selected to be one of the three projects of the European Space Agency's long-term plan, addressing the scientific theme of the Gravitational Universe. \citep{2013arXiv1305.5720E}. 
It consists of three spacecraft trailing the Earth around the Sun in a triangular configuration, with a mutual separation between spacecraft pairs of about 2.5 million kilometers, as shown by the cartoon representation of Figure \ref{fig:lisa_decigo}, left panel. The laser beams that connect the three satellites combine via time delay interferometry. The whole system is equivalent to pair of Michelson interferometers operating as a network. Because of its long arm length, LISA will be most sensitive in the millihertz frequency regime. The proposed launch year for LISA is 2037, and the mission lifetime is four years, with a possible six-year extension. A test mission - called LISA pathfinder \citep{2016PhRvL.116w1101A,2018PhRvL.120f1101A} - was launched in 2015 to test the technology necessary for LISA. The goal of the LISA pathfinder was to demonstrate a noise level 10 times worse than needed for LISA, but it exceeded this goal by a large margin, approaching the LISA requirement noise levels.
LISA's scientific goals are numerous since the instrument sensitivity window is extremely rich in gravitational-wave sources. Among them, we mention studying the formation and the evolution of compact binary stars in our galaxy; tracing the origin, growth, and merger history of massive black holes across cosmic ages; probing the dynamics of dense nuclear clusters using extreme mass ratio inspirals; understanding the astrophysics of stellar black holes; exploring the fundamental nature of gravity; measuring the rate of expansion of the Universe; understanding stochastic gravitational wave backgrounds and their implications for the early Universe; searching for gravitational-wave bursts and unforeseen sources.

\begin{figure*}
\plotone{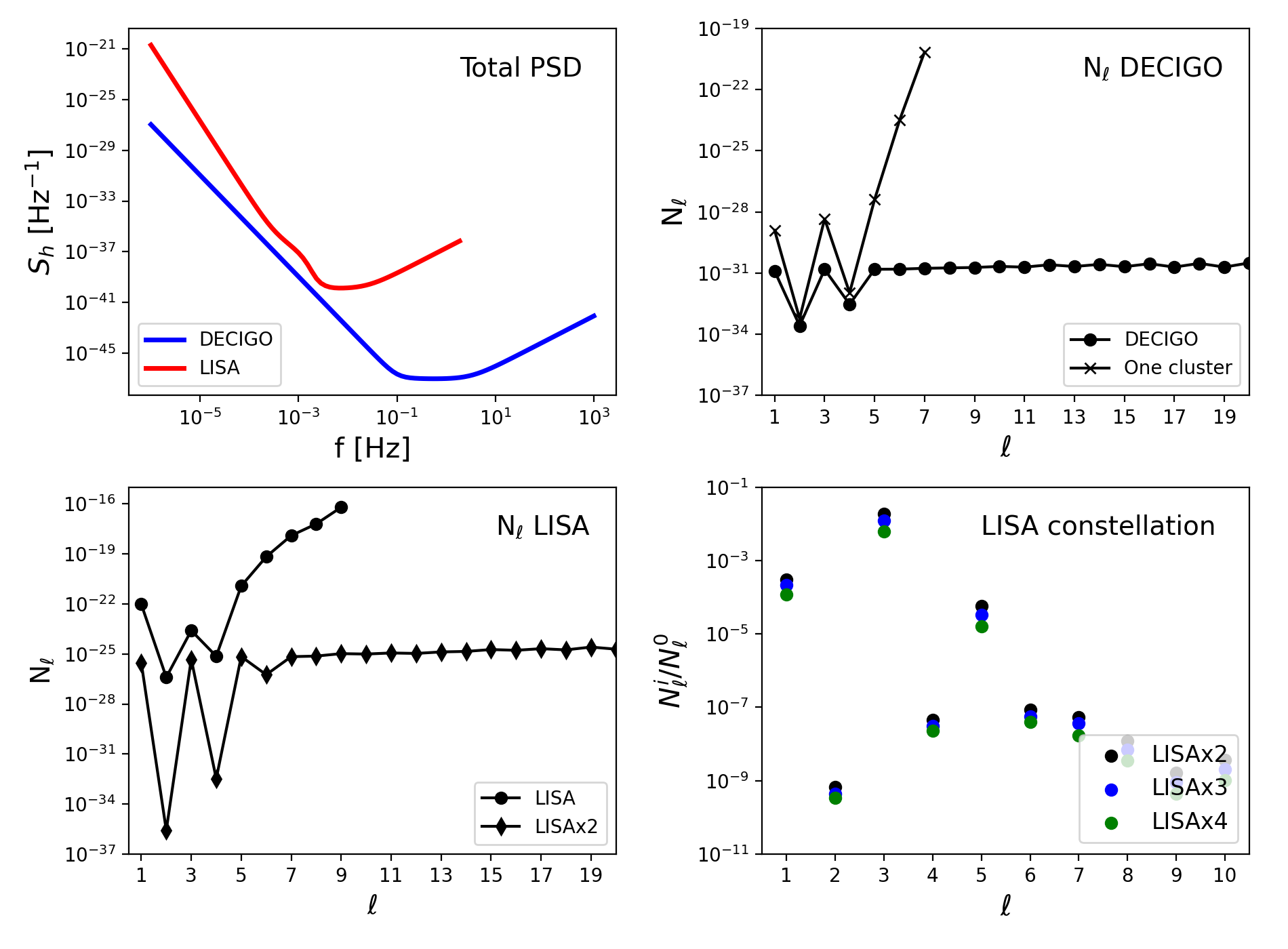}
\caption{\textit{Upper left panel}: LISA's and DECIGO's sensitivity curves expressed as total power spectral density. \textit{Upper right panel}: noise angular power spectra for DECIGO (four clusters) and a single cluster.
\textit{Lower left panel}: noise angular power spectra for LISA and constellation of two LISA-like instruments in orbit around the Sun. \textit{Lower right panel}: relative improvement of the $N_\ell$s of a constellation of LISA-like detectors with respect to LISA's ones for the various multipoles. 
\label{fig:angular_sensitivity}}
\end{figure*}

For this work, we calculate the LISA sensitivity curve using the parametric expression reported in \cite{2019CQGra..36j5011R}, assuming the nominal mission lifetime of four years. To compute the noise angular power spectrum $N_\ell$, we use the public code \texttt{schNell} \footnote{Publicly available at \url{https://github.com/damonge/schNell}.}, developed to calculate the angular power spectrum of the instrumental noise in interferometer networks mapping the SGWB \citep{2020PhRvD.101l4048A}. The code already contains LISA's specifications to compute its angular sensitivity. Moreover, we also modified the code to evaluate the angular sensitivity of a network of multiple LISA-like observatories in orbit around the Sun, spaced apart at equal distances along the orbit. We depict the specific configurations we consider for this work in the left panel of Figure \ref{fig:lisa_decigo}. Multiple LISA-like clusters operating together as a detector network have a better angular sensitivity because of the sensibly increased distance among detectors that constitutes a larger interferometric baseline. In fact, according to the Rayleigh criterion, the angular resolution of a GW detector is proportional to $\delta \theta \propto \lambda D / \rho$, where $\lambda$ is the GW wavelength, $\rho$ is its signal-to-noise ratio, and D is the effective size of the aperture (see \cite{2019arXiv190811410B} for example). A large aperture can be synthesized by having more than one detector operating simultaneously with a separation of a significant fraction of an astronomic unit, as is the case for the configurations shown in Figure \ref{fig:lisa_decigo}. In the lower right panel of Figure \ref{fig:angular_sensitivity}, we plot the ratio between the $N_\ell$s for different constellations of LISA-like clusters and the LISA's ones. We find that using two LISA-like clusters improves the angular sensitivity quite remarkably, especially for the even multipoles favored by the parity of LISA's antenna pattern. The angular sensitivity improves further by adding more detectors, even though the highest contribution results from adding the second cluster. Indeed, when we pass from one to two LISA-like clusters, the effective aperture of the network increases from a few million km to around two astronomic units. Instead, the aperture remains of the same order of magnitude when we add more than two clusters around the orbit. For this reason, there is a spectacular sensitivity improvement when adding a second LISA cluster and a modest one when adding more. Consequently, in this paper, we report only the analysis with a constellation with two LISA-like clusters, for simplicity.

\vspace{.5em}
\subsection{DECIGO}

DECIGO \citep {2021PTEP.2021eA105K,2017JPhCS.840a2010S,2011CQGra..28i4011K} is the planned Japanese space gravitational-wave antenna. It targets gravitational waves produced by astrophysical and cosmological sources in the 0.1 - 10 Hz frequency range. In this sense, DECIGO aims to bridge the frequency gap between LISA and ground-based detectors. A key advantage of DECIGO specializing in this frequency band is that the expected confusion noise, caused by irresolvable gravitational wave signals such as the ones from galactic white dwarf binaries, is low above 0.1 Hz. Moreover, DECIGO can serve as a follow-up for LISA by observing inspiraling sources that have moved above the mHz band or as a predictor for ground-based instruments by detecting inspiraling sources that have not yet moved into the 10 Hz - kHz band. 
DECIGO consists of four clusters of observatories in heliocentric orbit: two are in the same position, whereas the other two are evenly distributed around the Sun, as shown in the right panel of Figure \ref{fig:lisa_decigo}. Each cluster consists of three spacecraft, which form an equilateral triangle with a side of around 1000 km (the exact values of the parameters are still debated). Each instrument has a drag-free system and contains two mirrors floating inside the satellite as proof masses. DECIGO measures the change in the distance caused by between the two mirrors by employing a Fabry–Pérot cavity. Once launched, the mission lifetime of DECIGO will be at least three years. Before that, the DECIGO working group plans to launch the scientific pathfinder B-DECIGO in the 2030s to demonstrate the technologies required for DECIGO.
The most relevant goal for DECIGO is to detect primordial gravitational waves. However, there are many other scientific targets, such as probing the acceleration and expansion of the Universe, testing the accuracy of general relativity, examining the symmetry between the two polarizations of gravitational waves, and determining whether primordial black holes are a contributor to dark matter. Finally, DECIGO could reveal the formation mechanism of (super)massive black holes in the center of galaxies by detecting gravitational waves coming from intermediate mass ratio black hole binaries. Of course, the present work has relevance concerning this latter scientific objective. For more information about DECIGO's scientific targets, we remand the interested reader to \cite{2021PTEP.2021eA105K}. 

For this work, we computed DECIGO's sensitivity curve, including radiation pressure noise, shot noise, internal thermal noise, and gas thermal noise, following the prescriptions discussed in \cite{2021Galax...9...14I,2022Galax..10...25K}. In particular, we employed the detector parameters\footnote{Notice that the values of those parameters are not definitive and are still to be confirmed by the DECIGO working group.} presented in \cite{2022Galax..10...25K}. For the angular sensitivity, instead, we modified the public code \texttt{schNell} to compute the angular power spectrum of the instrumental noise of DECIGO. Specifically, we included an additional class of objects apt to describe a single DECIGO cluster. We then obtained the complete detector as a network of four clusters oriented and positioned around the orbit as depicted in the right panel of Figure \ref{fig:lisa_decigo}. In the upper right panel of Figure \ref{fig:angular_sensitivity}, we show the noise angular power spectrum for a single DECIGO cluster and the complete DECIGO configuration (four clusters).

\vspace{1.5em}
\section{Sources} \label{sec:sources}

We choose two astrophysical sources of gravitational waves as our case studies. First, we consider the SGWB produced by merging double compact objects (DCOs) of stellar origin, namely binary black holes (BH-BH) and neutron stars (NS-NS). As we will see in more detail in Section \ref{sec:monopole}, these sources emit gravitational waves in a broad frequency band. Even if the bulk of the signal resides in the Hz-kHz interval, we expect the SGWB produced by merging compact binaries to constitute one of the most relevant components in the deci-Hz band. Second, we choose the SGWB produced during one of the possible formation scenarios for massive black hole seeds. This process envisages multiple mergers of stellar remnants that sink toward the galactic center dragged by gaseous dynamical friction. Since the chirp mass and the mass ratio of the binaries involved in this scenario span a broad range of values, the SGWB frequency spectrum is very extended and includes the sensitivity bands of both LISA and DECIGO.

\vspace{.5em}
\subsection{Merging Double Compact Objects}
We characterize the population of DCOs and compute their merger rates following the approach presented in \cite{2019ApJ...881..157B,2021ApJ...907..110B,2020JCAP...10..045S}. The authors of these works combine the results of stellar population synthesis codes \citep{2017MNRAS.470.4739S,2015MNRAS.451.4086S,2019MNRAS.485..889S,2018MNRAS.474.2937C,2019MNRAS.482.5012C} with different observationally derived prescriptions for the host galaxies. For our analysis, we adopt the merger rates computed using the empirical Star Formation Rate Function as galaxy statistics and the Fundamental Metallicity Relation to assign metallicity to galaxies, combined with the results of the \texttt{STARTRACK} simulations\footnote{Simulation data publicly available at \url{https://www.syntheticuniverse.org/}}, specifically the `reference' model in \cite{2018MNRAS.474.2937C,2019MNRAS.482.5012C} (see Figure 8 of \cite{2021ApJ...907..110B}). This state-of-the-art method to compute the merger rates presents a twofold benefit. On the one hand, the galactic part is entirely observational-based, not relying on the results of any cosmological simulation or semi-analytic framework. On the other hand, since it uses the Star Formation Rate Function as galaxy statistics, one can assess the contribution of galaxies with different properties to the overall DCO merger rate. Of course, such empirical approaches also have their downsides. Mainly, the observational uncertainties affect all the final predictions, especially at high redshift.
The resulting merger rates are in good agreement with the recent local determination by the LIGO/Virgo/KAGRA collaboration, as shown in \cite{2021ApJ...907..110B,2021JCAP...11..032C}.
The redshift distribution of the merger rates highly depends on the star formation history of the host galaxies. Most of the BH-BH events come from $z \sim 2-3$, whereas most of the NS-NS ones come from slightly lower redshifts, $z \lesssim 2$. The chirp mass dependence, instead, is mainly determined by the stellar prescriptions and the derived DCO mass function, which is largely uncertain in the high-mass regime, where different formation channels may enter into play, complicating the evolutionary scenario (see \cite{2022ApJ...924...56S} for an example). All in all, the particular features of the merger rates strongly depend on the adopted astrophysical prescriptions: we refer the interested reader to \cite{2021ApJ...907..110B} for a more in-depth treatment. 
The overall normalization of the merger rates results from many different and complex physical processes related to stellar evolution that, in principle, could depend on the binary type (binary fraction, common envelope development/survival, natal kicks, mass transfers, etc.). In order to reduce the impact of the uncertainties in the astrophysical modeling, we decide to re-scale all the merger rates per unit comoving volume to match the values measured by LIGO/Virgo/KAGRA reported in \cite{2021arXiv211103634T}: $17.9 - 44\,\rm Gpc^{-3}\,yr^{-1}$ at $z = 0.2$ for BH-BH, and $10-1700\,\rm Gpc^{-3}\,yr^{-1}$ at $z=0$ for NS-NS, where the intervals are a union of 90 $\%$ credible intervals for the different methods used in the paper. Specifically, we calculate the logarithmic mean of the 90 $\%$ intervals and normalize our merger rates to retrieve those values at $z=0.2$ for BH-BH and $z=0$ for NS-NS. In this way, we maintain the redshift and chirp mass dependencies that we obtain with the methods described above but re-scale all the results to match the measured local values. Indeed, such a normalization directly affects all the results presented in this paper. Nonetheless, we stress that the local values of the merger rates we get with our calculations are inside the error bars of the LIGO/Virgo/KAGRA estimates, as it is shown in Figure 3 of \cite{2021JCAP...11..032C}.

\vspace{.5em}
\subsection{Massive Black Hole Seeds Formation}

The recent observations of high redshift quasars ($z \gtrsim 7$) powered by supermassive black holes (SMBHs) with $M > 10^{9} M_{\odot}$ (see \cite{2006AJ....131.1203F,2011Natur.474..616M,2017ApJ...851L...8V,2019ApJ...874L..30V} for example) have created tension between the estimated age of the Universe at those redshifts and the typical timescales of SMBH growth. In fact, at $z \gtrsim 7$, the age of the Universe was shorter than $\lesssim 0.8$ Gyr, whereas the accretion timescale driven by the gas disk (Eddington-like) is $\gtrsim 0.75$ Gyr. There are two classes of possible solutions to relieve this tension. The first way out invokes super-Eddington accretion rates, whereas the second involves mechanisms able to rapidly produce heavier BH seeds ($M \gtrsim 10^{3}-10^{5} M_{\odot}$), reducing the time required to attain the final billion solar masses by standard Eddington accretion.
In \citep{2020ApJ...891...94B,2021JCAP...10..035B}, the authors submit a new scenario to form heavy BH seeds, alternative or at least complementary to the other mechanisms. Specifically, they propose that BH seeds grow in the inner, gas-rich regions of dusty star-forming galaxies via multiple mergers with stellar compact remnants that migrate toward the center because of gaseous dynamical friction. Indeed, the dynamical drag subtracts energy and angular momentum from the moving object, making it sink toward the galactic center. The process is particularly efficient in dusty star-forming galaxies because they feature high star formation rates and huge molecular gas reservoirs concentrated in a compact region of a few kiloparsecs. These conditions foster the efficient sinking of innumerable compact remnants toward the galactic nucleus via gaseous dynamical friction. In \cite{2020ApJ...891...94B} the authors demonstrate that this mechanism can grow heavy BH seeds of masses $10^{4}-10^{6} M_{\odot}$ within some $10^{7}$ yr, so possibly alleviating the problem of supermassive BH formation at high redshift. With an accurate modeling of the gas distribution and the dynamical friction force, the authors of \citep{2020ApJ...891...94B} derive a fitting formula for the dynamical friction timescale. Consequently, they exploit the expression for the dynamical friction timescale to compute the merging rate of compact remnants at different galactic ages, so evaluating the contribution of this process to the growth of the central supermassive BH seed.
Of course, the repeated mergers of stellar BHs with the central growing seed would produce gravitational-wave emission, whose detection could be a smoking gun test for this scenario. In particular, the superposition of the unresolved gravitational-wave events constitutes an SGWB that extends over a wide range of frequencies (see \cite{2021JCAP...10..035B} and Section \ref{sec:monopole} for more detail).

\vspace{1.5em}
\section{Detection prospects for the monopole} \label{sec:monopole}

Stochastic gravitational-wave backgrounds are usually described through the dimensionless energy density parameter:

\begin{equation}
\omegagw(\fobs, \eobs ) 
= \frac{1}{\rho_{c}} \frac{d^{3} \rho_{\rm{gw}}(\fobs, \eobs )}{d \ln \fobs \, d^{2} \omega_{\rm{o}}}
= \frac{8 \pi G \fobs}{3H_{0}^{2}c^{2}} \frac{d^{3} \rho_{\rm{gw}}(\fobs, \eobs )}{d \fobs \, d^{2} \omega_{\rm{o}}}\,,
\end{equation}
where $\rho_{c} = 3H_{0}^{2}c^{2}/8\pi G$ is the critical density and $\rho_{\rm{gw}}$ is the SGWB energy density at the observed frequency $\fobs$, arriving from a solid angle $\omega_{\rm{o}}$ centred on the observed direction $\eobs$. The energy density parameter can be split into an isotropic term $\baromega (\fobs)$ and a directional dependent term $\delta \omegagw (\fobs, \eobs)$:

\begin{equation}
\omegagw(\fobs, \eobs) =\frac{\baromega (\fobs)}{4 \pi} + \delta \omegagw(\fobs, \eobs)\,.
\end{equation}
The isotropic term (aka the monopole) is obtained by summing the contribution of all the events at various frequencies (see for example \cite{2021JCAP...11..032C, 2021JCAP...10..035B}): 

\begin{equation} 
\baromega(\fobs)= \frac{8 \pi G \fobs} {3 H_{0}^{3} c^{2}} \int \frac{dz}{(1+z) h(z)} \int d \mc \;
\frac{d^{2} \dot{N}}{dV d\mc } \int dq \, \frac{dp}{dq} (q | \mc, z) \frac{dE}{df} (f_{e}(\fobs,z) | \mc,q), 
\end{equation}
where $\mc$ is the chirp mass, $f_{e} = (1+z) \fobs$ is the source frequency, $d^{2} \dot{N} / dV d\mc$ is the intrinsic merger rate per unit comoving volume and chirp mass, which we compute following the prescriptions described in Section \ref{sec:sources}, $h(z) = [ \Omega_{M}(1+z)^{3} + 1 - \Omega_{M}]^{1/2}$ accounts for the dependence of the comoving volume on cosmology, and $dE/df (f_{e}(z) | \mc)$ is the energy spectrum of the signal emitted by a single binary \citep{2008PhRvD..77j4017A}. With the previous expression, we can evaluate the intensity of the total SGWB, given by the incoherent superposition of all the events, resolved and unresolved. It is also possible to compute the residual SGWB by filtering out the gravitational-wave signals that lie above the detection threshold and are individually resolvable. The residual SGWB has a different frequency dependence and a globally lower amplitude (see \cite{2021JCAP...11..032C} for stellar DCOs and \cite{2021JCAP...10..035B} for massive black hole seeds formation). In this work, however, we focus on the total background since it has a higher amplitude and is more likely to be detected. 

\begin{figure*} 
\plotone{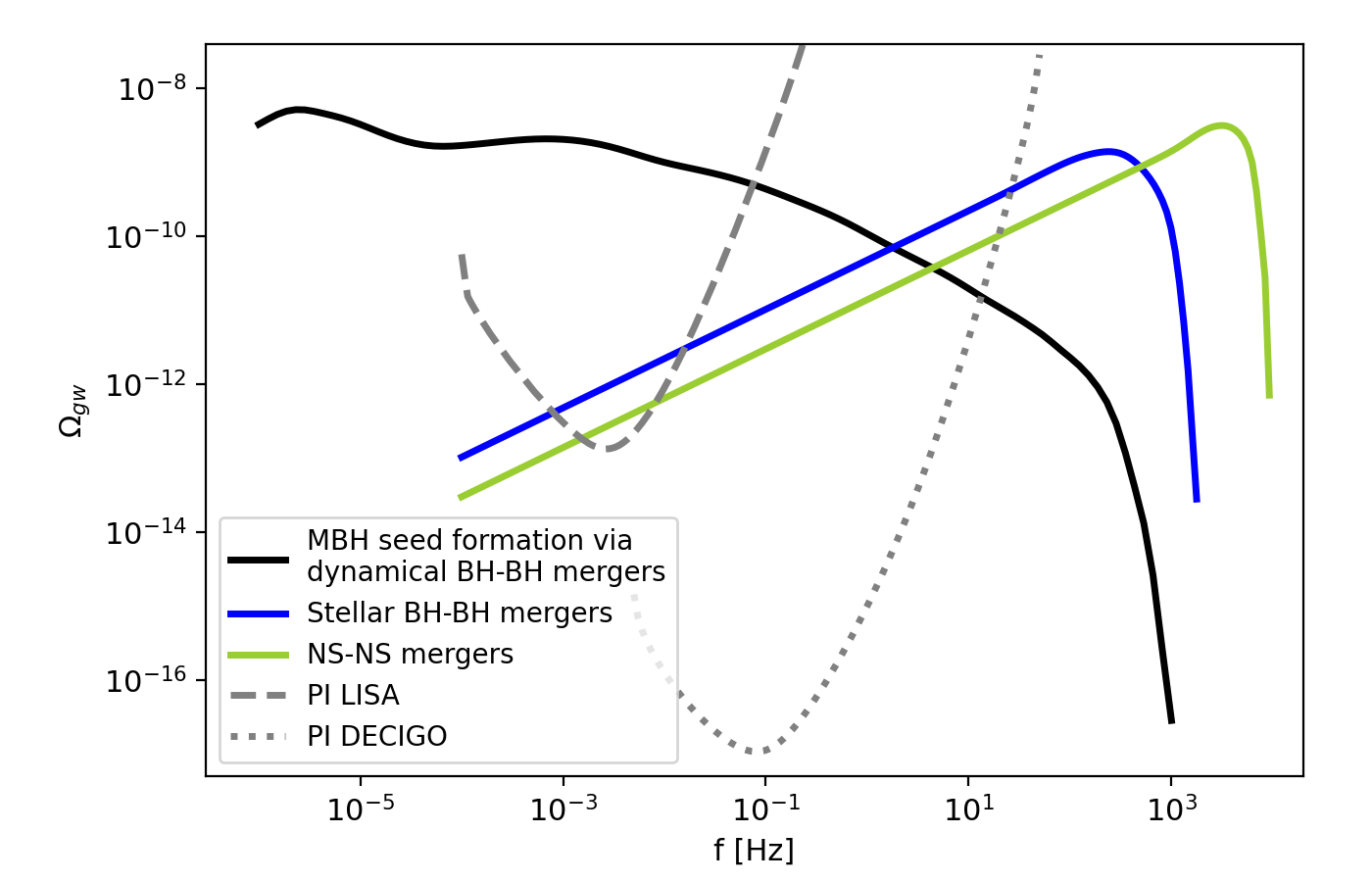}
\caption{The figure shows the frequency spectrum of the SGWB energy density for dynamical friction (black) and merging stellar black hole (blue) and neutron star (green) binaries. The grey lines represent the power law integrated sensitivity curves for LISA (dashed) and DECIGO (dotted), computed assuming total observing times of four and three years, respectively. \label{fig:omega_gw}} 
\end{figure*}

In figure \ref{fig:omega_gw}, we plot the isotropic energy density parameter of the two astrophysical SGWBs analyzed in this paper as a function of the observed frequency. The blue and green curves represent the monopole for merging stellar BH-BH and NS-NS binaries, respectively. At lower frequencies, as expected, the gravitational waves emitted during the inspiral phase dominate the signal, and the energy density parameter behaves as a power-law $\propto f^{2/3}$. At higher frequencies, the contribution from the merger phase becomes more and more relevant. After peaking at $f \sim 1$ kHz, the curves undergo an exponential drop related to the suppression of GW emission after the ringdown. The SGWB produced by merging DCOs has the largest amplitude in the frequency band of ground-based detectors. This signal will probably be observed for the first time at those frequencies, where it constitutes the dominant contribution to the total SGWB given by all possible sources. Still, the SGWB produced by stellar DCOs also has great relevance for the space-based interferometers that operate at lower frequencies. The dashed and dotted grey curves in figure \ref{fig:omega_gw} represent the power-law integrated (PI) sensitivity curves \citep{2013PhRvD..88l4032T} for LISA and DECIGO, respectively. The PI curve is a graphic representation of the detector sensitivity for SGWBs that considers the increase in sensitivity that comes from integrating over frequency other than time. This representation is strictly valid for SGWBs characterized by a power-law frequency dependence in the sensitivity band of the detectors, but it is usually employed to assess the ability to measure an SGWB of an instrument. Specifically, an SGWB whose energy density parameter is tangent to the PI curve has a signal-to-noise ratio equal to one. It follows that LISA should marginally detect the monopole of the SGWB produced by DCOs, whereas DECIGO will measure it with very high significance.
The black curve in figure \ref{fig:omega_gw} represents the frequency spectrum of the SGWB amplitude for the massive BH seeds formation process. The signal extends over a broad range of frequencies, including the sensitivity bands of both LISA and DECIGO, because the chirp masses of the involved binaries have very different values. In fact, at the beginning of the process, the central object is still very light, and the chirp mass assumes stellar values. As the central BH grows, the chirp mass increases and reaches values up to $10^{6}$ solar masses. High-mass mergers are more numerous and populate the low-frequency regime, whereas low-mass ones contribute at higher frequencies. In particular, the SGWB amplitude peaks around $10^{-6}-10^{-5}$ Hz, where there is a lack of planned gravitational-wave detectors\footnote{Actually, \cite{2022PhRvL.128j1103B,2022PhRvD.105f4021B} show that the $\mu$Hz gap could be filled by searching for deviations in the orbits of binary systems caused by their resonant interaction with gravitational waves.}. However, the amplitude remains more or less flat up to $10^{-1}$ Hz before experiencing a gradual decrease followed by an exponential drop at $f \gtrsim 10^{3}$ Hz, which corresponds to the lower possible chirp masses involved in the process. Comparing the SGWB monopole amplitude with the PI sensitivity curves, it follows that both LISA and DECIGO should be able to measure it with high significance. Because of its remarkable intensity in a frequency band where other SGWBs are less relevant, this signal - if detected - could be a smoking-gun probe of its origin process.

\vspace{1.5em}
\section{Detection prospects for the anisotropies} \label{sec:anisotropies}

\begin{figure*} 
\plotone{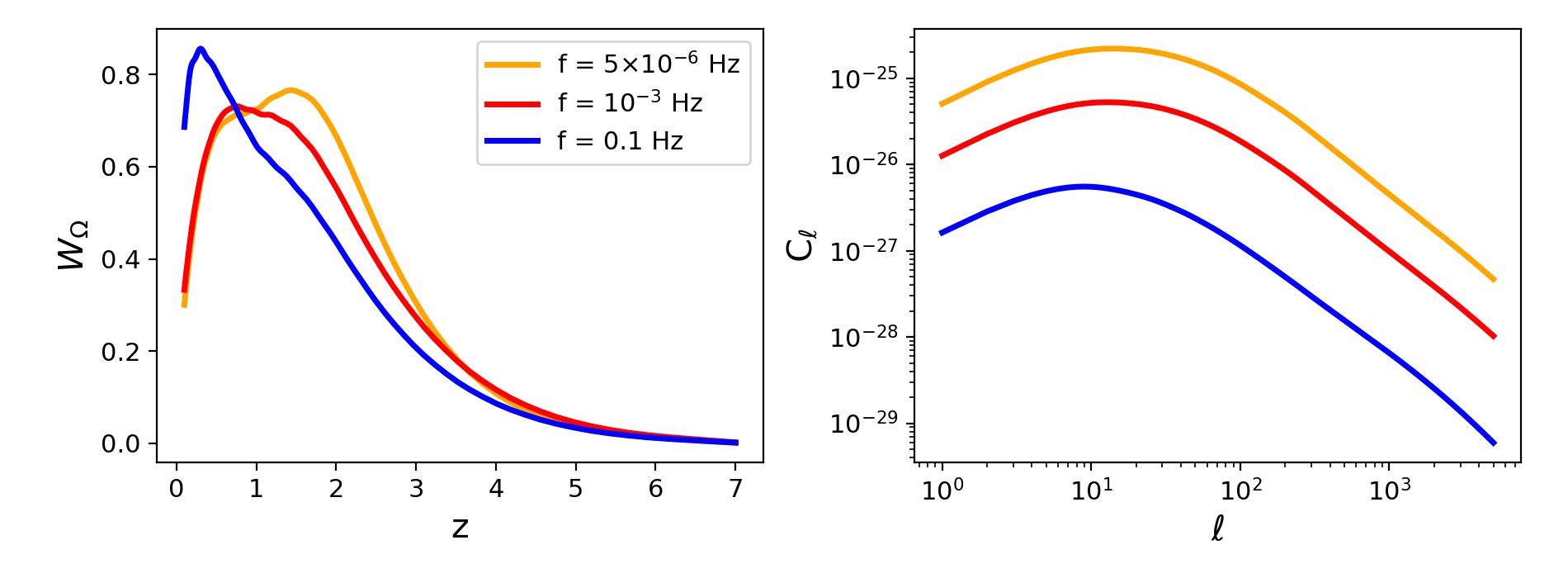}
\caption{\textit{Left panel}: Kernel of the SGWB produced by the massive BH seed formation process evaluated at three frequencies of interest. \textit{Right panel}: angular power spectrum of the anisotropies of the SGWB produced by the same process. \label{fig:kernel_cl_df}} 
\end{figure*}

\begin{figure*} 
\plotone{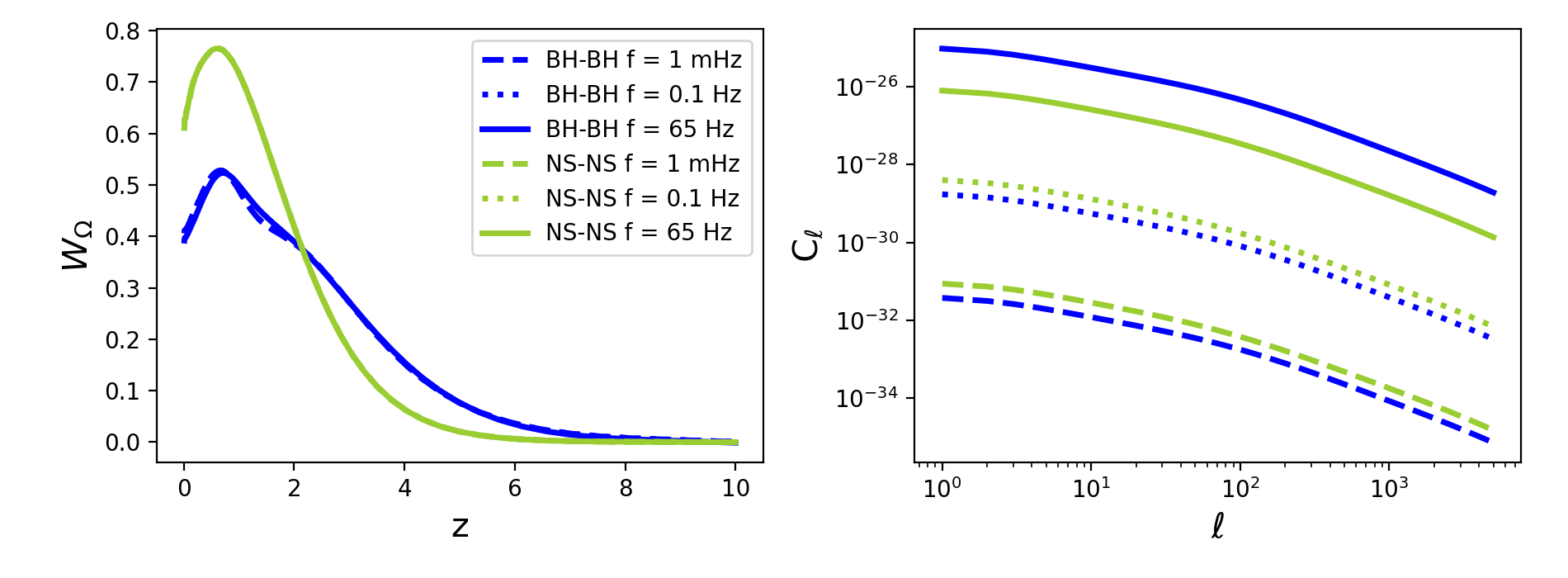}
\caption{\textit{Left panel}: Kernel of the SGWB produced by the merging BH-BH (blue) and NS-NS (green) binaries evaluated at three frequencies of interest. \textit{Right panel}: angular power spectrum of the anisotropies of the SGWB produced by merging compact binaries. \label{fig:kernel_cl_dco}} 
\end{figure*}

In this work, we calculate the angular power spectrum of the SGWB anisotropies according to the framework presented in \cite{2022Univ....8..160C}, which we briefly sketch hereafter. Assuming that the SGWB is a biased tracer of the underlying dark matter distribution, we can express the energy density contrast $\delta_{\rm{gw}} = \delta \omegagw / \baromega$ as a line-of-sight integral of the dark matter density contrast $\delta (\chi(z)\eobs, z)$:

\begin{equation}
  \delta_{\rm{gw}} (\fobs, \eobs) = \int_{0}^{z_{\star}} dz \; W^{\Omega}(\fobs,z) \delta (\chi(z)\eobs, z), 
\end{equation}
where $\chi (z)$ is the comoving distance to redshift z and $ z_{\star} = 1090$ is redshift at the last-scattering surface. The kernel $W^{\Omega}(\fobs,z)$ is the sum of two terms:

\begin{equation} \label{eq:sgwb_kernel}
  W^{\Omega}(\fobs,z) = \frac{b_{\Omega}(\fobs,z) \frac{d\Omega}{dz}(\fobs,z)}{ \biggl( \int dz' \frac{d\Omega}{dz'} \biggl) } + \mu(\fobs,z).
\end{equation}
The first term is the product of the linear bias $b_{\Omega}$, which quantifies the mismatch between the distribution of the SGWB and the total matter density, and the SGWB redshift distribution $d\Omega/dz$. We compute $b_{\Omega}$ and $d\Omega/dz$ as discussed in \cite{2021JCAP...11..032C,2022Univ....8..160C}. The second term accounts for the effect of weak gravitational lensing on the observed SGWB energy density. In this work, however, we consider only the total background that does not perceive any net impact from the weak lensing because its two effects - growth of energy density due to magnification and dilution of flux - balance each other in the absence of a detection threshold \citep{2020PhRvD.101j3513B,2021JCAP...11..032C}. For this reason, we will neglect the lensing term $\mu$ from now on. The left panels of Figures \ref{fig:kernel_cl_df} and \ref{fig:kernel_cl_dco} show the kernel of the SGWB produced by the massive BH seed formation mechanism and the merger of stellar DCOs, respectively, evaluated at various frequencies of interest. Even though the two processes are different and produce diverse SGWB signals (see Figure \ref{fig:omega_gw}), the kernels look similar because both the processes  - putting aside their intrinsic timescales - produce more GW events in concurrence with the peak of the cosmic star formation rate at $z \sim 2$. At higher redshifts, the kernels rapidly decrease because the binaries, and hence the GW events, are less and less numerous. Since the kernel $W_{\Omega}$ is a broad function of redshift for both the SGWB sources we consider for this work, we compute the angular power spectrum using the Limber approximation \citep{1954ApJ...119..655L} in the following way:

\begin{equation}
  C_{\ell}^{ \Omega} = \int _{0}^{z_{\star}} \frac{dz}{c} \frac{H(z)}{\chi^2(z)}\, \bigl[ W^{\Omega}(z) \bigl]^{2} \, P\biggl(k = \frac{l}{\chi(z)} ,z \biggl),
\end{equation}
where $c$ is the speed of light and $P (k, z)$ is the matter power spectrum, which we computed using the \texttt{CLASS}\footnote{The Cosmic Linear Anisotropy Solving System, available at \url{http://class-code.net}} public code \citep{2011arXiv1104.2932L,2011JCAP...07..034B}. We account for the nonlinear evolution of the matter power spectrum by using the \texttt{HALOFIT} prescription \citep{2003MNRAS.341.1311S}. 
The right panels of Figures \ref{fig:kernel_cl_df} and \ref{fig:kernel_cl_dco} show the angular power spectrum of the anisotropies for the SGWB produced by the two processes we are considering for this work. We evaluate the power spectra (and the kernels) at the frequencies $f = 1$ mHz and $f = 0.1$ Hz, optimal for a survey with LISA and DECIGO, and at the frequency where the two processes produce the signal with the largest amplitude. For massive black hole seed formation, this frequency is $f = 5\times10^{-6}$ Hz, whereas for merging compact binaries, we use $f = 65$ Hz.

\begin{figure*} 
\plotone{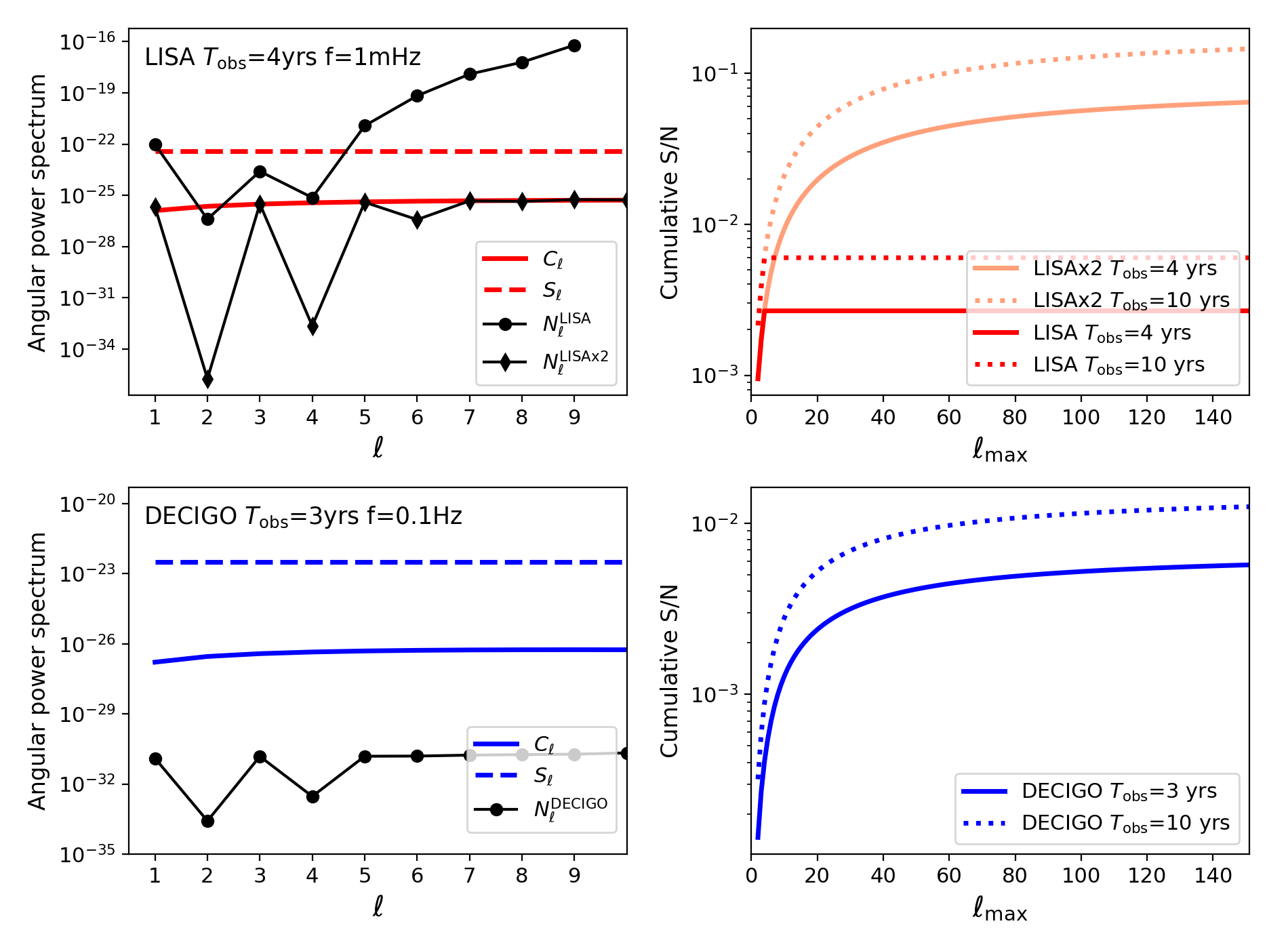}
\caption{In this figure, we present the results for the auto-correlation of the SGWB produced through the massive BH seed formation process. \textit{Upper left panel}: Angular power spectrum of the SGWB anisotropies, shot noise, and instrumental noise for four years of observation with LISA (or a constellation of two LISA clusters) at f = 1 mHz. 
\textit{Upper right panel}: Signal-to-noise ratio for the nominal LISA lifetime (four years, solid lines) and for an extended observation time of ten years (dashed lines). We show the results for LISA alone (red) and a constellation composed of two LISA-like clusters (orange). 
\textit{Lower left panel}: Angular power spectrum of the SGWB anisotropies, shot noise, and instrumental noise for three years of observation with DECIGO at f = 0.1 Hz. 
\textit{Lower right panel}: Cumulative signal-to-noise ratio for the nominal DECIGO lifetime (three years, solid lines) and for an extended observation time of ten years (dashed lines). 
\label{fig:results_auto_df}} 
\end{figure*}

\begin{figure*} 
\plotone{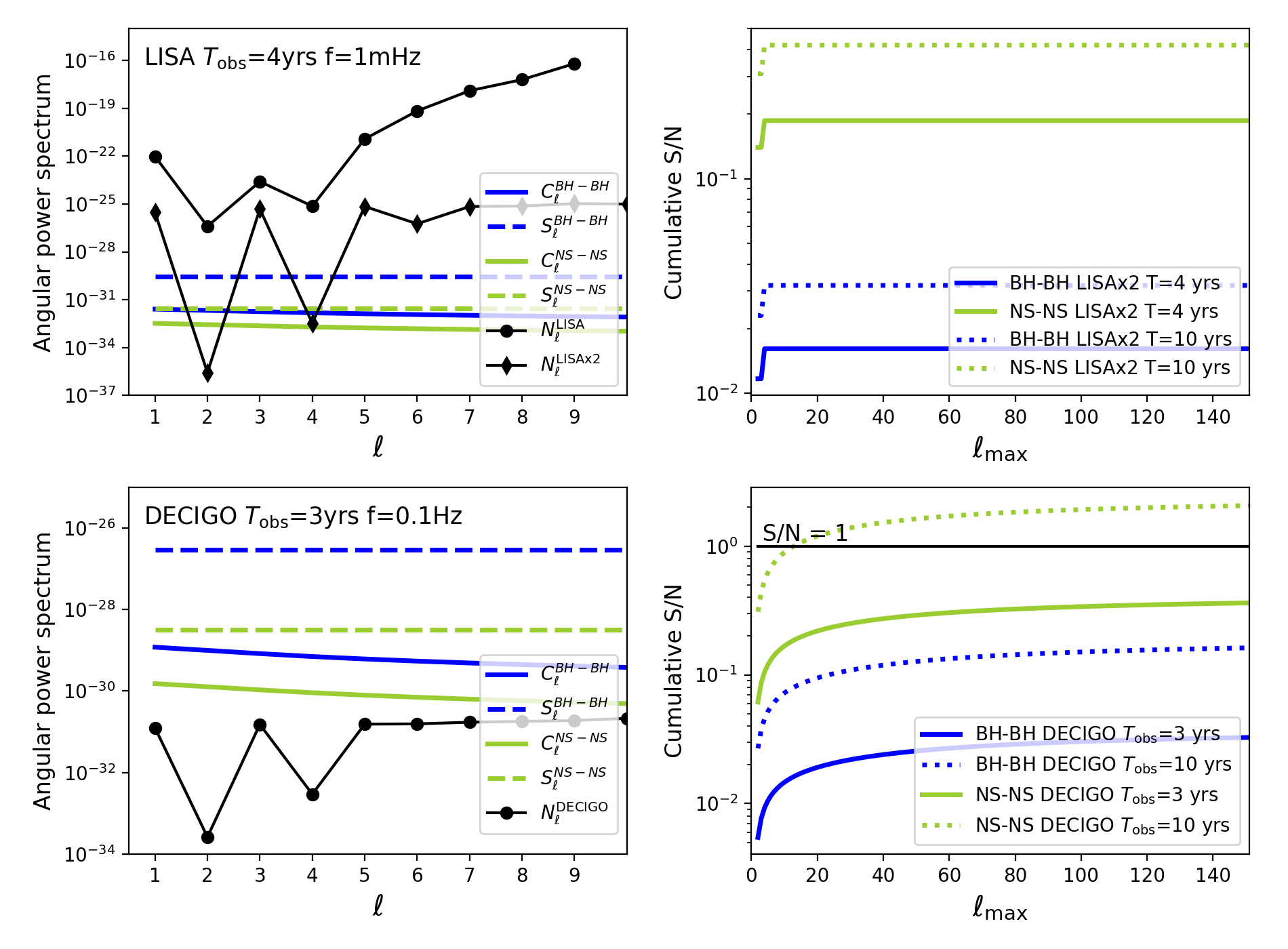}
\caption{In this figure, we present the results for the auto-correlation of the SGWB produced by merging compact binaries, specifically BH-BH (blue) and NS-NS (green). \textit{Upper left panel}: Angular power spectrum of the SGWB anisotropies, shot noise, and instrumental noise for four years of observation with LISA (or a constellation of two LISA clusters) at f = 1 mHz. 
\textit{Upper right panel}: Signal-to-noise ratio for the nominal LISA lifetime (four years, solid lines) and for an extended observation time of ten years (dashed lines). We only show the results for a constellation composed of two LISA-like clusters. 
\textit{Lower left panel}: Angular power spectrum of the SGWB anisotropies, shot noise, and instrumental noise for three years of observation with DECIGO at f = 0.1 Hz. 
\textit{Lower right panel}: Cumulative signal-to-noise ratio for the nominal DECIGO lifetime (three years, solid lines) and for an extended observation time of ten years (dashed lines).
\label{fig:results_auto_dco}} 
\end{figure*}

Assuming that the SGWB behaves as a Gaussian random field, the signal-to-noise ratio (S/N) of the $C_{\ell}^{\Omega}$ is given by: 

\begin{equation} \label{eq:snr}
  \biggl( \frac{S}{N} \biggl)_{\ell}^{2} = \frac{(2\ell + 1)}{2} \frac{ \bigl( C_{\ell}^{ \Omega} \bigl)^{2}}{\bigl(C_{\ell}^{\Omega} + S_{\ell}^{\Omega} + N_{\ell}^{\Omega} \bigl)^{2}},
\end{equation}
where $S_{\ell}^{\Omega}$ and $N_{\ell}^{\Omega}$ are the shot noise and the instrumental noise, respectively. We evaluated the former as discussed in Appendix B of \cite{2022Univ....8..160C} by exploiting a map-making technique that includes Poisson statistics and clustering properties. For the latter, we used the prescriptions described in Section \ref{sec:detectors}. All in all, the three main ingredients for the computation of the S/N are the angular power spectrum of the SGWB anisotropies $C_{\ell}$ (aka the signal), the shot noise $S_{\ell}$ and the instrumental noise $N_{\ell}$. To visually compare their contributions, in the left panels of Figures \ref{fig:results_auto_df} and \ref{fig:results_auto_dco}, we plot these quantities at the various multipoles for our two sources (merging DCOs and massive black hole seeds formation) and for the three considered detector configurations (LISA, constellation of two LISA-like clusters and DECIGO).
Throughout this work, we will assess the potential detectability of the various signals by using the cumulative S/N for multipoles up to $\ell_{\rm{max}}$, which is given by: 

\begin{equation} \label{eq:cumulative_snr}
  \biggl( \frac{S}{N} \biggl) (\ell < \ell_{\rm{max}}) = \sqrt{\sum_{\ell = \ell_{\rm{min}}}^{\ell_{\rm{max}}} \biggl( \frac{S}{N} \biggl)_{\ell}^{2}}.
\end{equation}
We show the results for the various cases in the right panels of Figures \ref{fig:results_auto_df} and \ref{fig:results_auto_dco}. For the SGWB produced during the formation of massive black hole seeds, it turns out that the auto-correlation of the anisotropies is not detectable in both LISA's and DECIGO's frequency bands. Around $f = 1$ mHz, both the shot noise and the LISA instrumental noise are too high compared to the angular power spectrum of the anisotropies, even if we consider the optimistic scenario of two LISA-like clusters operating for ten years. At DECIGO's frequencies, instead, the signal is higher than the instrumental noise, but the shot noise constitutes a killing factor for the overall S/N. Such a high shot noise depends on the small number of merger events contributing at $f = 0.1$ Hz. Indeed, the merger events that contribute to those high frequencies have a relatively low chirp mass ($\mc \lesssim 10^{3} M_{\odot}$) and are a tiny fraction of the total (see Figure 3 in \cite{2021JCAP...10..035B}).
The results for the SGWB produced by merging DCOs are only slightly better. A general comment is that the anisotropies of the SGWB produced by BH-BH are more intense than the ones for NS-NS because the overall amplitude of the SGWB is higher. Nevertheless, the shot noise is sensibly lower for binary neutron stars than for BH-BH since the NS-NS merger rate is at least two orders of magnitude higher. Consequently, a detector will measure much more NS-NS mergers during a given observation time, leading to a lower shot noise contribution. All in all, the S/N for NS-NS is higher than the S/N for BH-BH. As we already commented in Section \ref{sec:monopole}, in both LISA's and DECIGO's frequency range, the SGWB produced by merging DCOs has a power-law behavior $\propto f^{2/3}$. Therefore, at $f = 1$ mHz, the signal is too low to be detected even by two LISA-like clusters. Instead, at $f = 0.1$ Hz, the SGWB is slightly more intense: this, together with the incredibly low DECIGO's instrumental noise, causes the S/N to be higher. In particular, for an observation time $T = 10$ yrs, the S/N for NS-NS reaches the threshold value of one for $\ell_{\rm{max}} \sim 10$.

\vspace{1.5em}
\section{Cross-correlation with CMB lensing} \label{sec:cross}

The take-home message from the previous Section is that a measure of the SGWB anisotropies is almost unattainable with a constellation of space-based interferometers. The combined effects of instrumental and shot noise make the undertaking extremely arduous, even with a long observation time. As already suggested in many works, a possible way to enhance the intrinsic SGWB anisotropies is the cross-correlation with another tracer of the Large-Scale Structure. Existing papers discuss the benefits of cross-correlating the astrophysical SGWB with galaxy number counts \citep{2020PhRvD.102b3002A,2020PhRvD.102d3513C,2021MNRAS.500.1666Y,2020MNRAS.491.4690M}, weak lensing \citep{2017PhRvD..96j3019C,2019PhRvD.100f3004C}, and CMB temperature and polarization fluctuations  \citep{2021PhRvL.127A1301R,2021PhRvD.104l3547B,2022Univ....8..160C}. Here, we re-propose the cross-correlation with the CMB lensing convergence. We characterize CMB lensing as a tracer of the Large-Scale Structure following the approach presented in \cite{2022Univ....8..160C}, which - in turn - is inspired to \cite{2015ApJ...802...64B,2016ApJ...825...24B}.
The CMB lensing convergence $\kappa$ is defined as the laplacian of the lensing potential $\phi$, and we can express it as a weighted integral over redshift of the projected dark matter density contrast $\delta$:

\begin{equation}
  \kappa (\eobs) = - \frac{1}{2} \nabla^{2} \phi (\eobs) = \int_{0}^{z_{*}} dz \; W^{\kappa}(z) \delta (\chi(z)\eobs, z),
\end{equation}
where $\chi (z)$ is the comoving distance to redshift z, and $ z_{\star} = 1090$ is the redshift at the last-scattering surface. The weight inside the integral is the lensing kernel $W^{\kappa}$, which describes the lensing efficiency of the matter distribution and is given by

\begin{equation} \label{eq:lensing_kernel}
  W^{\kappa}(z) = \frac{3 \Omega_{m}}{2c} \frac{H_{0}^{2}}{H(z)} (1+z) \chi(z) \frac{\chi_{*} - \chi(z)}{\chi_{\star}}, 
\end{equation} 
where $\chi_{\star}$ is the comoving distance to the last-scattering surface and $\Omega_{m}$ and $H_{0}$ are the present-day value of the matter density and the Hubble parameter, respectively. 
Similarly to what we did for the auto-correlation, we compute the angular power spectrum of the cross-correlation as: 

\begin{equation}
  C_{\ell}^{\kappa \Omega} = \int _{0}^{z_{\star}} \frac{dz}{c} \frac{H(z)}{\chi^2(z)}\, W^{\kappa} (z) W^{\Omega}(z) \, P\biggl(k = \frac{l}{\chi(z)} ,z \biggl).
\end{equation}
Assuming that also the CMB lensing is a Gaussian field, the S/N of the cross-correlation is given by:

\begin{equation}
  \biggl( \frac{S}{N} \biggl)_{\ell}^{2} = \frac{(2\ell + 1) f_{\rm{sky}}\bigl( C_{\ell}^{\kappa \Omega} \bigl)^{2}}{\bigl(C_{\ell}^{\kappa \Omega}\bigl)^{2} + \bigl(C_{\ell}^{\kappa} + N_{\ell}^{ \kappa} \bigl)\bigl(C_{\ell}^{ \Omega} + S_{\ell}^{\Omega} + N_{\ell}^{ \Omega} \bigl)}.
\end{equation}
In the previous expression, $f_{\rm{sky}}$ is the sky fraction covered by both the SGWB and the CMB surveys, $C_{\ell}^{\kappa}$ is the auto-correlation angular power spectrum of the CMB lensing convergence and $N_{\ell}$ is the lensing noise. Notice that in the lensing-related terms in the denominator (aka the cross-correlation and the lensing convergence auto-correlation), we have only the contribution from cosmic variance since the shot noise is absent when considering a diffuse field such as the lensing convergence. For our analysis, we employ the lensing noise curves for CMB-S4 \citep{2019arXiv190704473A,2022arXiv220308024A} and the Simons Observatory \citep{2019JCAP...02..056A}. For both surveys, we adopt the Large Aperture Telescope configuration sky fraction $f^{\kappa}_{\rm{sky}} = 0.4$. Since gravitational-wave experiments cover the entire sky (i.e. $f^{\Omega}_{\rm{sky}} = 1$), we use the limiting value $f_{\rm{sky}} = f^{\kappa}_{\rm{sky}}$ for the cross-correlation. In the following, we will show only the results for CMB-S4 for a matter of simplicity. Indeed, it has lower noise, and the final S/N does not strongly depend on the adopted lensing noise curve. Moreover, the Simons Observatory will be active in a few years, while CMB-S4 is likely to be more contemporary to the GW detectors we are considering for this work.

\begin{figure*} 
\plotone{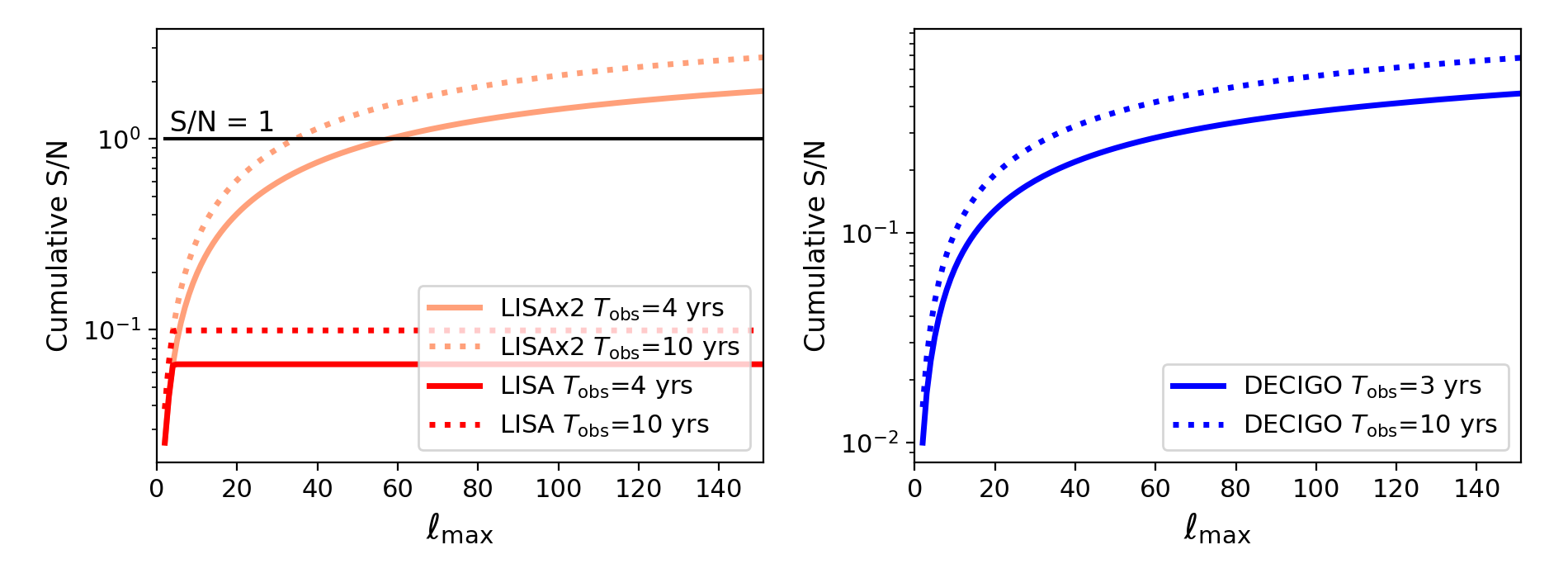}
\caption{Cumulative signal-to-noise ratio of the cross-correlation between the SGWB produced by massive BH seed formation process and the CMB lensing convergence, measured with CMB-S4. \textit{Left panel}: S/N for LISA (red) and a constellation of two LISA-like clusters (orange) observing for four (solid) and ten (dotted) years. The black line represents the detection threshold corresponding to S/N = 1.
\textit{Right panel}: S/N for DECIGO observing for three (solid) and ten (dotted) years. 
\label{fig:results_xc_df}} 
\end{figure*}

\begin{figure*} 
\plotone{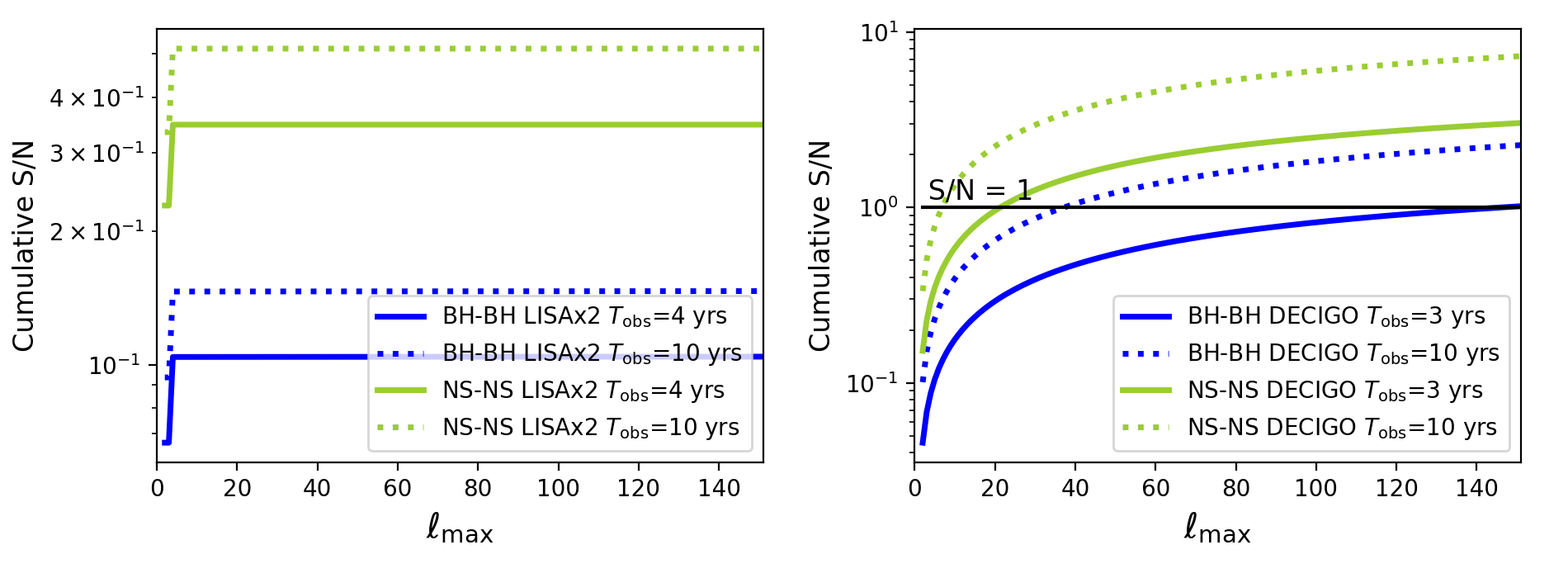}
\caption{Cumulative signal-to-noise ratio of the cross-correlation between the SGWB produced by merging compact binaries (BH-BH in blue and NS-NS in green) and the CMB lensing convergence, measured with CMB-S4. \textit{Left panel}: S/N for a constellation of two LISA-like clusters with four (solid) and ten (dotted) years of observing time. 
\textit{Right panel}: S/N for DECIGO observing for three (solid) and ten (dotted) years. The black line represents the detection threshold corresponding to S/N = 1.
\label{fig:results_xc_dco}} 
\end{figure*}

In Figure \ref{fig:results_xc_df}, we show the S/N of the cross-correlation angular power spectrum for the SGWB produced during the formation of massive black hole seeds. As discussed in the previous Section, the shot noise in the DECIGO band is too high and compromises the S/N even if we cross-correlate the SGWB with CMB lensing. In the mHz band, instead, the situation is more promising. As expected, LISA alone is not able to resolve the signal, but with a constellation of two clusters\footnote{We performed the analysis also for the constellations of three and four LISA-like clusters depicted in Figure \ref{fig:lisa_decigo}. The results improve only slightly (less than $1\%$) with respect to the two-cluster configuration. In fact, the $N_{\ell}$s are only a few times lower (see Figure \ref{fig:angular_sensitivity}) and, in any case, smaller than the $C_{\ell}$s and $S_{\ell}$s evaluated at the multipoles that contribute most to the S/N ($\ell = 2$ and $\ell = 4$, see the left panels of Figures \ref{fig:results_auto_df} and \ref{fig:results_auto_dco}).}, the cross-correlation S/N reaches unity at $\ell_{\rm{max}} \sim 50$ for $T_{\rm{obs}} = 4$ yrs and at $\ell_{\rm{max}} \sim 30$ for $T_{\rm{obs}} = 10$ yrs. In principle, this means that one should be able to probe the signal by summing the contributions from enough multipoles. However, we still don't know if the large noise contributions will allow us to obtain the $C_{\ell}$ from the raw maps up to such high $\ell_{\rm{max}}$. 
In Figure \ref{fig:results_xc_dco}, we show the results for the cross-correlation between the CMB lensing convergence and the SGWB produced by merging DCOs. In the LISA band, the SGWB amplitude is too low to access the signal, even with the help of cross-correlations. In the DECIGO band, instead, there are more chances to measure the cross-correlation signal. For the nominal mission lifetime of $T_{\rm{obs}} = 3$ yrs, the S/N reaches unity at $\ell_{\rm{max}} \sim 20$ for NS-NS, while it always stays below one for BH-BH. However, if we consider a longer observation time $T_{\rm{obs}} = 10$ yrs, the S/N will reach unity at $\ell_{\rm{max}} \lesssim 10$ for NS-NS and $\ell_{\rm{max}} \lesssim 40$ for BH-BH. 

\vspace{1.5em}
\section{Conclusion} \label{sec:conclusion}
In this paper, we investigated the possibility of probing the anisotropies of the astrophysical SGWB by using constellations of space-based interferometers. For our analysis, we considered a network composed of multiple LISA-like clusters and the planned Japanese mission DECIGO, which is already a constellation of four detector clusters. We tested these detector configurations with two different anisotropic SGWBs of astrophysical origin. First, we considered the SGWB produced by merging stellar compact binaries: even though this signal is dominant in the Hz-kHz band, its contribution is also relevant in the mHz and deci-Hz bands. Second, we considered the SGWB produced during a newly proposed scenario for massive black hole seed formation by consecutive mergers of stellar remnants brought into the galactic center by the gaseous dynamical friction. This process, recently developed in \citep{2020ApJ...891...94B,2021JCAP...10..035B}, produces an SGWB in an extended frequency band, ranging from $10^{-7}$ Hz up to 1 Hz, making it a potential target for space-based interferometers.
Following the formalism presented in \citep{2022Univ....8..160C}, we computed the angular power spectrum of the SGWB anisotropies for both sources, the shot noise, and the instrumental noise for all the considered detector configurations. We then used these ingredients to evaluate the signal-to-noise ratio for the auto-correlation power spectra. As expected, we found that measuring the SGWB anisotropies is mostly unattainable with a constellation of space-based interferometers. A possible exception is for the SGWB produced by merging binary neutron stars, observed with DECIGO for at least ten years. In this case, the cumulative S/N reaches unity at $\ell_{\max} \lesssim 10$, at least with our prescriptions. Of course, there are conspicuous uncertainties in the astrophysical modeling, and the exact amplitude of the signal could be very different from the one we computed from the present data.
The prospects sensibly improve when considering cross-correlation with other tracers of the Large-Scale Structure. In particular, we cross-correlated our SGWB signals with the CMB lensing convergence. We computed the S/N of the cross-correlation power spectrum by including the lensing noise curves for CMB-S4 and the Simons Observatory. We found that a constellation of two LISA-like clusters operating for ten years can marginally probe the cross-correlation between CMB lensing and the SGWB produced during the formation of massive black hole seeds. Specifically, the S/N reaches unity at $\ell_{\rm{max}} \sim 30$. Moreover, we found that DECIGO can instead probe the cross-correlation between CMB lensing and the SGWB produced by merging compact binaries. For an observation time of ten years, the S/N reaches unity at $\ell_{\rm{max}} \lesssim 10$ for binary neutron stars and $\ell_{\rm{max}} \lesssim 40$ for binary black holes. 
All in all, the anisotropies of the SGWB contain a richness of astrophysical and cosmological information that makes them a sought-after target for present and future gravitational-wave observatories. However, many observational difficulties will probably prevent probing them with ground-based interferometers. With this preliminary analysis, we showed that using a constellation of space-based interferometers could improve the angular sensitivity enough to probe the anisotropies of the SGWB, at least through cross-correlations with other cosmic fields, such as the CMB lensing.

\begin{acknowledgments}
We thank Yuki Kawasaki, Seiji Kawamura, and Tomohiro Ishikawa for fruitful discussions about DECIGO's expected noise modeling. Specifically, their help was fundamental to computing DECIGO's noise correlation matrix, which is necessary to evaluate the $N_{\ell}$. We warmly thank Giulia Cusin for carefully reading the manuscript and for useful discussions. Lastly, we thank the anonymous referee for the valuable comments that helped us improve the quality of the paper. AL acknowledges funding from the EU H2020-MSCA-ITN-2019 Project 860744 \textit{BiD4BESt: Big Data applications for black hole Evolution STudies} and the PRIN MIUR 2017 prot. 20173ML3WW, \textit{Opening the ALMA window on the cosmic evolution of gas, stars, and supermassive black holes}. GC, CB, and LB acknowledge partial support by the INDARK INFN grant. CB acknowledges support from the COSMOS $\&$ LiteBIRD Networks by the Italian Space Agency (\url{http://cosmosnet.it}) 
\end{acknowledgments}




\bibliography{bib_constellation}
\bibliographystyle{aasjournal}

\end{document}